\documentclass[a4paper,11pt]{article}
\usepackage{jcappub}
\usepackage{graphicx}

\subheader{Accepted by JCAP}

\title{Statistics of Dark Matter Halos in the Excursion Set Peak Framework}
\author[a,b]{A. Lapi}
\author[b]{L. Danese}
\affiliation[a]{Dip. Fisica, Univ. `Tor Vergata', Via Ricerca Scientifica 1, 00133 Roma, Italy}
\affiliation[b]{SISSA, Via Bonomea 265, 34136 Trieste, Italy}
\emailAdd{lapi@sissa.it}
\emailAdd{danese@sissa.it}

\abstract{We derive approximated, yet very accurate analytical expressions for the abundance and
clustering properties of dark matter halos in the excursion set peak framework;
the latter relies on the standard excursion set approach, but also includes the
effects of a realistic filtering of the density field, a mass-dependent threshold for
collapse, and the prescription from peak theory that halos tend to form
around density maxima. We find that our approximations work excellently for
diverse power spectra, collapse thresholds and density filters. Moreover,
when adopting a cold dark matter power spectra, a tophat filtering and a
mass-dependent collapse threshold (supplemented with conceivable scatter), our
approximated halo mass function and halo bias represent very well the outcomes of
cosmological $N-$body simulations.}


\begin{document}
\maketitle
\flushbottom

\section{Introduction}

According to the standard paradigm, galaxies and galaxy systems form when
baryons settle within the gravitational potential wells constituted by
virialized dark matter (DM) `halos'. To deeply understand the abundance and
clustering properties of these halos is a fundamental step toward formulating
a sensible theory for the formation and evolution of cosmic structures in the
Universe.

The issue is very complex, and its attack ultimately requires brute-force
$N-$body simulations (see
\cite{sheth01,springel05,warren06,reed07,tinker08,murray13,watson13});
however, some analytic grasp is essential to physically interpret their
outcomes, to provide approximated yet flexible analytic representations of
the results, to develop strategies for future setups, and to quickly explore
the effect of modifying the background cosmological/cosmogonical model.

In this vein, two main frameworks have been developed: the excursion set
approach (see
\cite{press74,epstein83,bond91,sheth02,maggiore10,paranjape12,corasaniti11,musso12,lapi13}),
and peaks theory (see \cite{bardeen86,appel90,bond96,manrique98}). Basically,
the former statistically describes the mass fraction in the density field
which is above a critical threshold for collapse when conveniently smoothed
on a certain scale. The latter envisages that halos form around special
positions in space, and specifically around the peaks of the density field.
Actually, the prescriptions from the two approaches can be merged in a
unified framework, constituting a sort of excursion set theory for peaks, or
excursion set peaks in brief (see
\cite{appel90,jedamzik95,nagashima01,paranjape12}). Recently, it has been
shown that, with reasonable assumptions on the collapse history of halos,
such a framework can be extremely effective in reproducing the outcomes of
cosmological $N-$body simulations (see \cite{paranjape13a}).

Unfortunately, the results of the excursion set peak framework cannot be put
in closed analytic form; this would be particularly useful to inspire
consistent fitting formulae of the simulation outcomes, to provide flexible
descriptions for different background cosmologies, and to easily derive other
statistical quantities of interest, like halo merger rates. The main goal of
this paper is to present approximated, yet very accurate analytical
expressions for the halo abundance and bias in the excursion set peak
framework.

Our working plan is straightforward: in section 2 we recall the basics of the
excursion set approach, of peaks theory, and of their connection in the
excursion set peak framework; in section 3 we present our analytic
approximation to the excursion set peak, compare it to the exact results and
to the outcomes of cosmological $N-$body simulations; in section 4 we
summarize our findings.

As to cosmology we adopt the standard, flat Universe (see \cite{planck13})
with matter density parameter $\Omega_M=0.32$, baryon density parameter
$\Omega_b=0.05$, and Hubble constant $H_0 = 100\, h$ km s$^{-1}$ Mpc$^{-1}$
with $h=0.67$. As to cosmogony, we adopt the standard cold DM power spectrum
$P(k)$ by \cite{bardeen86} with the correction for baryons by
\cite{sugiyama95}, normalized in such a way that the r.m.s. $\sigma(R)\equiv
\sqrt{S(R)}$ takes on the value $\sigma_8=0.82$ on a scale $R=8\, h^{-1}$
Mpc.

\section{The excursion set peak framework}

In this Section we recall the basics of the excursion set peak framework,
highlight some delicate points useful in the sequel, and set the notation.
The expert reader may jump directly to section 3.

We consider a Gaussian initial field of fluctuations with overdensity
$\delta\equiv \rho/\bar\rho-1$ relative to the background $\bar \rho$; we
indicate with $\delta_R$ the overdensity field smoothed on some scale $R$
through a filter function $W_R(r)$ enclosing a volume $V=4\pi\int{\rm
d}r~r^2\, W_R(r)\propto R^3$. The variance of this field
\begin{equation}
S_R\equiv \langle\delta_R^2\rangle=\int{{\rm d}k\over 2\pi^2}~k^2\,
P(k)\,W_R^2(k)~,
\end{equation}
is determined by the Fourier transform of the filter $W_R(k)$ and by the
power spectrum $P(k)$; for the spectra of interest in cosmology, $S_R$ is an
inverse, monotonic function of $R$.

The excursion set approach bases on the notion that, as the smoothing scale
$R$ decreases from large values, $\delta_R$ executes a random walk as a
function of the increasing variable $S_R$ (e.g.,
\cite{bond91,maggiore10,lapi13}); an example is illustrated in figure 1. The
theory envisages that a halo with mass $M=\bar \rho\, V$ is formed when the
walk first crosses a barrier $B(S,t)$ with general shape
\begin{equation}
B(S,t)= \delta_c(t)\,\left[1+\beta\,\left({S\over
\delta_c^2}\right)^\eta\right] = \nu\,(1+\beta\,\nu^{-2\eta})~.
\end{equation}
Here $\delta_c(t)$ is the critical threshold for collapse extrapolated from
linear perturbation theory; at the current epoch $\delta_c\approx 1.686$
holds, although the precise value weakly depends on cosmological parameters,
and then it evolves like $\delta_c(t)\propto 1/D(t)$ with the cosmological
time $t$, in terms of the linear growth function $D(t)$, see \cite{eke96} for
details.

The dependence of the barrier on the scale $S$ is specified by the parameters
$(\beta,\eta)$, that are commonly set basing on the classic theory of halo
collapse and/or on comparison with numerical simulation. The value $\beta=0$
corresponds to a constant barrier that describes the standard spherical
collapse (\cite{press74,bond91}). On the other hand, numerical simulations
indicate the collapse is rather ellipsoidal (if not triaxial);
correspondingly, the barrier features a nonlinear shape with typical values
$\beta\approx 0.5$ and $\eta\approx 0.45-0.65$. Note that the constraints on
the slope $\eta$ are rather loose, and the value $\eta=0.5$ marking a
`square-root barrier' is often adopted because of its simplicity (cf. section
2.1), and to avoid delicate normalization issues (for $\eta>1/2$ not all
walks are guaranteed to cross the barrier since the r.m.s. height scales like
$\sqrt{S}$). Actually, simulations also show that the barrier $B(S,t)$ at
given $S$ features some scatter, whose ultimate nature is currently unknown
(see \cite{sheth01,robertson09,despali13}). The last equality in eq. (2.2)
highlights that these barriers can be recast in self-similar terms with the
use of the variable $\nu\equiv \delta_c(t)/\sqrt{S}$; this is useful to treat
simultaneously different epochs and/or masses\footnote{Note that in the
literature on the excursion set approach, including our paper \cite{lapi13},
the variable $\nu$ is often defined as $\delta_c^2/S$; here instead we adopt
the standard notation of the excursion set peak framework and define
$\nu\equiv \delta_c/\sqrt{S}$.}.

According to the excursion set prescription, the halo mass function, i.e. the
number density of halos of given mass $M$, is given by
\begin{equation}
{{\rm d}N\over {\rm d}M} = {\bar \rho\over M^2}\, \left|{{\rm d log}\nu\over
{\rm d log}M}\right|\, \nu\,f(\nu)~;
\end{equation}
here $f(S)=f(\nu)\,|{\rm d}\nu/{\rm d}S|=\nu\,f(\nu)/2\,S$ is the first
crossing distribution, i.e., $f(S)\, {\rm d}S$ represents the probability
that a trajectory crosses the barrier for the first time between $S$ and
$S+{\rm d}S$. Since there is a one-to-one correspondence between the halo
mass function and the first crossing distribution, in the rest of the paper
we will present our results in terms of the latter.

As to the filter function, the simplest choice is a tophat filter in Fourier
space $W_R(k)\propto \theta_{\rm H}(k_R-k)$, with $\theta_{\rm H}(\cdot)$ the
Heaviside step function and $k_R\propto 1/R$ a cutoff wavenumber; the
resulting random walk is purely Markovian, with no correlation between the
steps. Such a `sharp $k-$space filter' have been adopted at large in the past
literature since it allows to derive analytically exact expressions of the
mass function for simple barrier shapes (e.g., constant, square-root or
linear). For other nonlinear barriers the result is only numerical (e.g.,
\cite{zhang06,zentner07,benson13}) but can be approximated analytically (see
\cite{lapi13}) in the mass range relevant for cosmological studies. However,
the outcome with this filter is somewhat unsatisfactory, mainly for two
reasons: (i) the filter is unrealistic since the enclosed volume $V$ (hence
the mass $M$) is ill-defined or, at least, ambiguous; (ii) a general
agreement with the mass function from cosmological simulations can be
obtained only after rescaling the nonlinear barrier of eq. (2.2) by an ad-hoc
factor $\sqrt{q}$ with $q\approx 0.7$.

Admittedly, a tophat filter in real space $W_R(r)\propto \theta_{\rm H}(R-r)$
offers a more direct contact with the standard theory of nonlinear halo
collapse, and a fairer comparison with numerical simulations since better
reflects the way halos are identified into them\footnote{As shown by
\cite{tinker08} and many other authors, the first crossing distribution
extracted from $N-$body simulations depends on the way halos are identified;
in this work we always refer to the $N-$body outcome for halos identified via
a spherical overdensity algorithm with a nonlinear contrast $\Delta\approx
200$. This can be fairly compared with the excursion set results for a
real-space top-hat filtering and a linear density contrast $\delta_c\approx
1.686$.}. The corresponding Fourier transform of the tophat filter which
enters eq. (2.1) reads
\begin{equation}
W_R(k)=3\,{\sin k R-k R\cos k R\over (k R)^3}~;
\end{equation}
this filter encloses a finite and well-defined volume $V=4\pi\, R^3/3$ ,
although for some power spectra it can produce a smoothed density field which
is not locally differentiable with respect to $r$. To circumvent the issue,
it is standard to multiply the above expression by a small-scale Gaussian
$e^{-\epsilon\, k^2/2}$ with $\epsilon\lesssim 0.1$, that smooths the sharp
edges of the tophat (see \cite{bond91,maggiore10,musso12}). Another
filtering, which is commonly used in peaks theory, is constituted by a pure
Gaussian function $W_R(k)=e^{k^2\, R^2/2}$, which plainly produces a
differentiable smoothed density field and encloses a volume $V=(2\pi)^{3/2}\,
R^3$.

With the tophat and gaussian filter, the random walk executed by $\delta_R$
as a function of $S_R$ is no longer Markovian, and correlations between the
steps are introduced. The quantity
\begin{equation}
C_{R,R'}\equiv \langle\delta_R\delta_{R'}\rangle=\int{{\rm d}k\over
2\pi^2}~k^2\, P(k)\, W_{R}(k)\, W_{R'}(k)~,
\end{equation}
gauges the strength of these correlations; the outcome is that random walks
with correlations tends to vary less erratically than the Markovian one, as
can be seen in the examples illustrated in figure 1.

Moreover, the numerical and analytic techniques to derive the first crossing
distribution in the Markovian case cannot be applied when correlations
between the steps are present, and other kinds of computations must be
employed.

\subsection{Upcrossing in place of first crossing}

A Montecarlo approach is the most direct way to derive the first crossing
distribution (see \cite{bond91,farahi13}); this exploits the fact that the
probability of a trajectory to reach a location $(\delta_{R+\Delta
R},S_{R+\Delta R})$ starting from $(\delta_R,S_R)$ is simply given by a
gaussian with mean $C_{R,R+\Delta R}\, \delta_R/S_R$ and variance
$[S_{R+\Delta R}-C_{R,R+\Delta R}^2/S_R]^{1/2}$. The Montecarlo algorithms
runs as follows: a set (typically $10^5$ realizations are enough) of walks is
constructed by extracting randomly the increment in $\delta_R$ at each step
in $S_R$ from the above distribution; then each walk is evolved until it
first crosses the barrier at a given location $S$; the number of walks
crossing the barrier between $S$ and $S+\Delta S$, divided by $\Delta S$ and
by the number of overall realization, is the first crossing distribution
$f(S)$. Such a Montecarlo scheme has the advantage of providing the
\emph{exact} (i.e., not approximated) first crossing distribution, although
the outcome is only numerical, and the required computational time may become
long when massive use of $f(S)$ is needed.

To circumvent the issue, some analytic approximations have been developed in
the literature, aimed at reproducing the outcomes of the Montecarlo
computation above. A simple and rather accurate one has been originally
proposed by \cite{bond91} for a constant barrier, and then extended to
nonlinear barriers by \cite{musso12}. The idea is to approximate the `first
crossing' distribution $f(S)$ with the `upcrossing' one, i.e., the
distribution of walks crossing the barrier from below, irrespective of
earlier crossings. The bottom line is that when correlations between the
steps are present, the random walk does not exhibit the frenetic variations
of the Markovian case; thus if it crosses the barrier from below at $S$, it
is very likely that no crossing has occurred for $S'<S$, and especially so at
small $S$.

Quantitatively, one has to compute the fraction of walks with the
requirements that $\delta_R> B(S_R)$ and $\delta_{R+\Delta R} < B(S_{R+\Delta
R})$ for small $\Delta R$. This translates into $f(S) \simeq \int_{\dot
B}^{\infty}{\rm d}\dot\delta~(\dot \delta-\dot B)\, p(B,\dot\delta)$, where
$p(B,\dot\delta)$ is the joint distribution of $\delta=B$ and $\dot
\delta\equiv {\rm d}\delta/{\rm d}S$. Given that by the definitions eqs.
(2.2) and (2.5) one has $\langle\delta^2\rangle=S$ and
$\langle\delta\dot\delta\rangle=1/2$ (cf. also appendix A), the joint
distribution $p(B,\dot\delta)$ is a bivariate that can be written as the
product of two Gaussians, one $p(B)$ with zero mean and variance $S$, and the
other $p(\dot\delta|B)$ with mean
$B\,\langle\delta\dot\delta\rangle/\langle\delta^2\rangle=B/2S$ and variance
$\langle\dot\delta^2\rangle-\langle\delta\dot\delta\rangle^2/\langle\delta^2
\rangle=(1-\gamma^2)/4 \gamma^2\,S$, in terms of the parameter $\gamma\equiv
1/\sqrt{4\,S\,\langle\dot\delta^2\rangle}$.

Transforming from $\dot \delta$ to $\xi\equiv
2\gamma\,\sqrt{S}\,(\dot\delta-\dot B)$, the outcome writes
\begin{eqnarray}
\nonumber \nu\,f(\nu)&\simeq& {e^{-B^2/2\,S}\over
\sqrt{2\pi}\,\gamma}~\int_0^\infty{\rm d}\xi~\xi\,
{e^{-(\xi-w)^2/2\,(1-\gamma^2)}\over
\sqrt{2\pi\,(1-\gamma^2)}}=\\
\\
\nonumber &=&{w\,e^{-B^2/2\,S}\over \sqrt{2\pi}\,\gamma}
\,\left\{{1\over 2}\,{\rm erfc}\left[-{w\over
\sqrt{2\,(1-\gamma^2)}}\right]+\sqrt{1-\gamma^2\over
2\pi}\,{e^{-w^2/2\,(1-\gamma^2)}\over w}\right\}~,
\end{eqnarray}
in terms of the scaling variable $\nu$ of eq. (2.2) and of the quantity
$w\equiv -2\,\gamma S\, {\rm d}[B(S)/\sqrt{S}]/{\rm d}S =
\gamma\nu\,[1+(1-2\,\eta)\,\beta\,\nu^{-2\,\eta}]$; for a constant
($\beta=0$) or a square-root ($\eta=1/2$) barrier, the latter just reads
$w=\gamma\, \nu$. In these cases, the first crossing distribution can be put
in the particularly simple form
\begin{equation}
\nu\,f(\nu)\simeq {\nu\,e^{-(\nu+\beta)^2/2}\over
\sqrt{2\pi}}\,\left\{{1\over 2}\,{\rm erfc}\left[-{\Gamma\nu\over
\sqrt{2}}\right]+{e^{-\Gamma^2\,\nu^2/2}\over \sqrt{2\pi}\,\Gamma \nu}\right\}~,
\end{equation}
with $\Gamma\equiv \gamma/\sqrt{1-\gamma^2}$. As shown in appendix A the
quantity $\Gamma$ is a constant for a scale-invariant spectrum, but for a
cold DM spectrum it instead depends on the scale $\nu$ (i.e., it depends on
the scale $S$ at given redshift, or equivalently on redshift at given scale
$S$); this marks a break of self-similarity in the first crossing
distribution, that actually appears to be indicated by some recent numerical
simulations, and is at variance with what happens in the uncorrelated case.
However, the dependence $\Gamma(\nu)$ is mild (see appendix A), hence to a
first approximation it can be neglected and the value at $\nu\approx 1$ can
be effectively adopted. Specifically, $\Gamma\approx 0.85$ and $0.51$ apply
for the Gaussian and the tophat filter, respectively.

In figure 2 we illustrate by the red lines how the upcrossing distribution of
eq. (2.6) approximates very well the first crossing distribution derived by
the Montecarlo method, both for a constant and a square-root barrier. We also
plot, for comparison, other approximations that have been devised in the
literature: the green line refers to the result for a constant barrier by
\cite{maggiore10}, which is based on a perturbative path integral approach to
the excursion set; the magenta lines illustrate the approximations by
\cite{farahi13}, which may be thought as the nonperturbative version of the
path integral approach; the blue lines illustrate the result for complete
correlation by \cite{paranjape12}, which considers deterministic walks with
heights increasing monotonically as $\delta_R\propto \sqrt{S_R}$, and
corresponds to the limit $\gamma\simeq 1$ in eq. (2.6); the cyan lines refer
to the approximation by \cite{peacock90}, which assumes the walk can be
broken into independent segments with only internal correlations; the yellow
lines illustrate for reference the result for a sharp $k-$space filter,
corresponding to uncorrelated (Markovian) walks. In the bottom panel, the
inset shows the residuals of these approximations with respect to the exact,
Montecarlo outcome.

All in all, the approximation provided by the upcrossing distribution is by
far the best compromise between simplicity and accurateness; e.g., for the
square-root barrier the relative deviation amounts to $\Delta \log
\nu\,f(\nu)\lesssim 0.1$ for $-1.5\lesssim\log \nu^2\lesssim 1$. We note,
however, that when compared to the outcome from cosmological $N-$body
simulations by \cite{tinker08}, the first crossing distribution from
excursion set theory, either exact or approximated, strongly underpredicts
the number density of large mass objects (small $S$, large $\nu^2$), and
features a very different overall shape; actually it performs even worse than
the uncorrelated case. A possible solution to such a shortcoming is discussed
next.

\subsection{Peaks as special positions}

In the standard formulation of the excursion set theory, one assumes that
halos may form equally likely at every point in space. However, $N-$body
simulations suggest that the formation events occur preferentially around
maxima, i.e. peaks, of the density field. The classic peaks theory tells us
how to perform a weighted average of the walks over such special positions in
the underlying field (see
\cite{bardeen86,appel90,bond96,nagashima01,paranjape12,paranjape13b}).

In peaks theory, the variable $x\equiv 2\gamma\sqrt{S}\,\dot\delta = \xi +
2\gamma\sqrt{S}\,\dot B\equiv \xi+\tau$ underlying eq. (2.6) basically
represents the curvature of the density field (see \cite{appel90}, their
section 5.2); this is strictly true only for Gaussian smoothing, but it holds
to a good approximation also for tophat filtering. Thus an excursion set
theory for peaks, or excursion set peaks, may be obtained just by accounting
for the distribution of curvatures $F(x)$ around a peak position. Such a
distribution is given by \cite{bardeen86} (see their eq. A15)
\begin{eqnarray}
\nonumber F(x) &=& {1\over 2}\,(x^3-3\,x)\,\left[{\rm erf}\left(x\sqrt{5\over
2}\right)+{\rm erf}\left(x\sqrt{5\over 8}\right)\right]+\\
\\
\nonumber &+& {2\over 5\pi}\,\left[\left({31\over 4}\,x^2+{8\over 5}\right)\,e^{-5\,x^2/8}+
\left({1\over 2}\,x^2-{8\over 5}\right)\,e^{-5\,x^2/2}\right]~,
\end{eqnarray}
in terms of a function quite rapidly increasing with $x$; asymptotically, the
expansions $F(x)\simeq 3^5\, 5^{3/2}\, x^8/7\times 2^{11}\,\sqrt{2\pi}$ for
$x\ll 1$ and $F(x)\simeq x^3-3\, x$ for $x\gg 1$ apply.

Correspondingly, the first crossing distribution for the excursion set peak
framework writes (see \cite{paranjape13a})
\begin{equation}
\nu\,f(\nu)\simeq {e^{-B^2/2\,S}\over \sqrt{2\pi}\,\gamma}\, {V\over
V_\star}~\int_0^\infty{\rm d}\xi~\xi\, F(\xi+\tau)\,
{e^{-(\xi-w)^2/2\,(1-\gamma^2)}\over \sqrt{2\pi\,(1-\gamma^2)}}~,
\end{equation}
with $\tau\equiv 2\gamma\sqrt{S}\dot B =
2\,\eta\,\gamma\beta\,\nu^{1-2\,\eta}$; for a constant barrier $\tau=0$,
while for a square-root barrier $\tau=\gamma\beta$. In addition, $V/V_\star$
is a the ratio of the volume enclosed by the filter to a characteristic
volume defined by \cite{bardeen86}; such a ratio increases with $\nu$, as
detailed in appendix A.

Figure 3 illustrates that the net effect of including the peak constraints
yields considerably more objects at the high-mass end relative to the
upcrossing distribution, and an overall shape more similar to the $N-$body
outcomes, especially for the square-root barrier. We will come back to the
comparison of the excursion set peak results with numerical simulations at
the end of Sect. 3.

\section{Approximated excursion set peak}

Now we aim at obtaining useful, approximated analytic expressions of the
first crossing distribution for the excursion set peak framework. We start by
deriving the asymptotic expansion of eq. (2.9) in the limit of small $S\ll
\delta_c^2$ or $\nu\gg 1$, i.e., large masses and/or early times.

In this limit, $w\gg 1$ and the integral appearing in eq. (2.9) is dominated
by large values of $\xi\simeq w$. Then we note that the integrand comprises
two functions of $\xi$: a rapidly declining exponential, and the rapidly
increasing function $F$; their product is a bell-shaped function with a clear
maximum at a location $\bar \xi$, that suggests to try a Gaussian
approximation. In other words we require in a neighborhood of $\xi\simeq \bar
\xi$ that
\begin{equation}
F(\xi+\tau)\,e^{-(\xi-w)^2/2\,(1-\gamma^2)}\simeq F(\bar \xi+\tau)\,e^{-(\bar
\xi-w)^2/2\,(1-\gamma^2)-\kappa\, (\xi-\bar \xi)^2/2}~,
\end{equation}
with $\bar \xi$ and $\kappa$ two quantities to be determined. At the zeroth
order in $\xi$ this expression is plainly true.

Requiring that it also holds at the first order implies
\begin{equation}
{{\rm d}\over {\rm d}\xi}
\left[F(\xi+\tau)\,e^{-(\xi-w)^2/2\,(1-\gamma^2)}\right]_{|\,{\bar \xi}}=0~.
\end{equation}
Since $\xi\simeq \bar \xi$ is supposed to be large, we can use the expansion
$F(x)\simeq x^3-3\, x$; the result is the equation
\begin{equation}
3\,(1-\gamma^2)\,[(\bar\xi+\tau)^2-1] =
(\bar\xi-w)\,(\bar\xi+\tau)\,[(\bar\xi+\tau)^2-3]
\end{equation}
which yields the asymptotic expansion
\begin{equation}
\bar \xi\simeq w\left[1+{3\,(1-\gamma^2)\over w^2}\right]~.
\end{equation}

Requiring that eq. (3.1) holds even at the second order implies
\begin{equation}
{{\rm d^2}\over {\rm d}\xi^2}
\left[F(\xi+\tau)\,e^{-(\xi-w)^2/2\,(1-\gamma^2)}\right]_{|\,{\bar \xi}}=
-\kappa\,F(\bar \xi+\tau)\, e^{-(\bar \xi-w)^2/2\,(1-\gamma^2)}~;
\end{equation}
the result is the equation
\begin{equation}
6\,(\bar\xi+\tau)-6\,[(\bar\xi+\tau)^2-1]\,{\bar \xi-w\over
1-\gamma^2}-{[(\bar\xi+\tau)^2-3]\,(\bar\xi+\tau)\over
1-\gamma^2}\,\left[1-{(\bar\xi-w)^2\over 1-\gamma^2}-\kappa\,(1-\gamma^2)\right]~,
\end{equation}
which after eq. (3.4) yields the asymptotic expansion
\begin{equation}
\kappa\simeq {1\over 1-\gamma^2}\,\left[1+{3\,(1-\gamma^2)\over w^2}\right]~.
\end{equation}

Now we can substitute in eq. (2.9) the gaussian approximation eq. (3.1) to
obtain
\begin{equation}
\nu\,f(\nu)\simeq {e^{-B^2/2\,S}\over \sqrt{2\pi}\,\gamma}\, {V\over
V_\star}~F(\bar \xi+\tau)\, {e^{-(\bar \xi-w)^2/2\,(1-\gamma^2)}\over
\sqrt{2\pi\,(1-\gamma^2)}}\,\int_0^\infty{\rm d}\xi~\xi\, e^{-\kappa\,(\xi-\bar
\xi)^2/2}~;
\end{equation}
the gaussian integral is trivial, while the function in front of it can be
expanded with the help of eqs. (3.4) and (3.6) to yield
\begin{eqnarray}
\nonumber \nu\,f(\nu)&\simeq& {w\,e^{-B^2/2\,S}\over \sqrt{2\pi}\,\gamma}\, {V\over
V_\star}~\left[w^3+3\,\tau\,w^2+3\,(1+\tau^2-2\,\gamma^2)\, w\right]\times \\
\\
\nonumber &\times & \left\{{1\over 2}\,{\rm
erfc}\left[-{w\over \sqrt{2\,(1-\gamma^2)}}\right]+\sqrt{1-\gamma^2\over
2\pi}\,{e^{-w^2/2\,(1-\gamma^2)}\over w}\right\}~;
\end{eqnarray}
remarkably, this is the same result of the upcrossing distribution eq. (2.6),
modulated by a function of $w$.

The approximation of eq. (3.9) strictly holds for large $w$, but can be
improved so as to work pretty well in the whole range relevant for
cosmological studies. To this purpose, note that when $w\ll 1$ then the
integral in eq. (2.9) tends to a non-null constant; in our asymptotic
expansion eq. (3.9) instead the modulating function in square brackets makes
everything to vanish in that limit. To recover the correct behavior for small
$w$, it is sufficient to add in the square brackets a constant term; we find
by trials that the value $(1+2\,\beta)\,(1-\gamma^2)$ works pretty well in
yielding the exact result in the limit $w=0$ for diverse values of $\gamma$
and $\beta$.

All in all, our approximation of the first crossing distribution for the
excursion set peaks framework writes
\begin{equation}
\nu\,f(\nu)\simeq  {w\,e^{-B^2/2\,S}\over \sqrt{2\pi}\,\gamma}\, {V\over
V_\star}~\mathcal{P}(w)\, \left\{{1\over 2}\,{\rm
erfc}\left[-{w\over \sqrt{2\,(1-\gamma^2)}}\right]+\sqrt{1-\gamma^2\over
2\pi}\,{e^{-w^2/2\,(1-\gamma^2)}\over w}\right\}~.
\end{equation}
where we have defined the polynomial $\mathcal{P}(w)\equiv
w^3+3\,\tau\,w^2+3\,(1+\tau^2-2\,\gamma^2)\, w+(1-\gamma^2)\,(1+2\,\beta)$.

For a constant or a square-root barrier with $w=\gamma\nu$ and
$\tau=\gamma\beta$, the previous equation can be put in the remarkably
simpler form
\begin{equation}
\nu\,f(\nu)\simeq {\nu\,e^{-(\nu+\beta)^2/2}\over \sqrt{2\pi}}\,{V\over
V_\star}\,\mathcal{P}(\gamma\nu)\,\left\{{1\over 2}\,{\rm
erfc}\left[-{\Gamma\nu\over \sqrt{2}}\right]+{e^{-\Gamma^2\,\nu^2/2}\over
\sqrt{2\pi}\,\Gamma \nu}\right\}~,
\end{equation}
in terms of $\Gamma=\gamma/\sqrt{1-\gamma^2}$. The novel expressions in eqs.
(3.10) and (3.11) constitute the main results of the present paper.

In figure 3 the comparison between the orange dotted and dashed lines
(actually superimposed at the high-mass end) illustrates how efficiently the
above expression approximate the exact result; e.g., for the square-root
barrier the typical deviation amounts to $\Delta\log \nu\,f(\nu)\approx 0.05$
for $-1.5\lesssim \log\nu^2\lesssim 1.5$. It is seen that our approximation
constitutes also a fair representation of the $N-$body outcome (see
\cite{tinker08}; crosses); the inset shows the corresponding residuals.
However, we notice that for a constant barrier the high-mass end is largely
overpredicted, while for the square-root barrier it is slightly
underpredicted. On the other hand, such a deficit of massive objects can be
easily offset on considering the scatter in the value of the parameter
$\beta$, as suggested by numerical simulations (see
\cite{sheth01,robertson09,despali13}). One can obtain the scattered first
crossing distribution $\tilde f(S)$ just by convolving the distribution at
fixed $\beta$ from our approximations eqs. (3.10) and (3.11), with the
distribution of $\beta$. For the sake of simplicity, we consider the outcome
eq. (3.11) for a square-root barrier and a Gaussian scatter with mean value
$\langle\beta\rangle = 0.45$ and variance $\Sigma_\beta=0.35$, consistent
with simulation indications. Since the dependence on $\beta$ in the
modulating function $\mathcal{P}(\gamma\nu,\gamma\beta)$ is mild, it can be
neglected in performing the convolution. Then the result is analytic and
reads
\begin{eqnarray}
\nonumber \nu\, \tilde f(\nu)&=&\int{\rm
d}\beta~{e^{-(\beta-\langle\beta\rangle)^2/2\Sigma_{\beta}^2}\over
\sqrt{2\pi\Sigma_{\beta}^2}}\,\nu\,f(\nu) =\\
\\
\nonumber & = & {\nu\,e^{-(\nu+\beta)^2/2\,(1+\Sigma_\beta^2)}\over
\sqrt{2\pi\,(1+\Sigma_\beta^2)}}\,{V\over
V_\star}\,\mathcal{P}(\gamma\nu)\,\left\{{1\over 2}\,{\rm erfc}\left[-{\Gamma\nu\over
\sqrt{2}}\right]+{e^{-\Gamma^2\,\nu^2/2}\over \sqrt{2\pi}\,\Gamma
\nu}\right\}~;
\end{eqnarray}
this expression constitutes a novel result. Note that for more complex
barriers (e.g., $\eta>1/2$) or $\beta-$distributions (e.g., lognormal) the
above convolution must be performed numerically, but the net effect is
similar to what we have found.

The result, illustrated in figure 3 (bottom panel) by the orange solid line,
reproduces very well the $N-$body outcome; it is seen from the residuals
plotted in the inset that the relative deviation amounts to $\Delta \log
\nu\,f(\nu)\lesssim 0.1$ for $-1.5\lesssim\log \nu^2\lesssim 1$. In terms of
the characteristic mass $M_\star(z)$ defined by
$\delta_c^2(z)=S[M_\star(z)]$, the range translates into $M\gtrsim 10^{-2}\,
M_\star$ at $z\approx 0$, $M\gtrsim 10^{-3}\, M_\star$ at $z\approx 1.5$, and
practically encompasses all masses of cosmological interest for $z\gtrsim
1.5$.

Finally, in figure 4 we show that our approximation of Eq.~(3.11) works
excellently also for filters different from a tophat, and for power spectra
different from a standard cold DM. To wit, we show the results for a Gaussian
and tophat filters when scale-invariant power spectra are used in place of
the standard cold DM one; a square-root barrier has been adopted for
definiteness.

\subsection{Halo bias}

To show a direct application of our above results on halo abundance, we now
focus now on deriving an approximated expression for the halo bias in the
excursion set peak framework. The bias is the classical way of quantifying
the connection between halo abundance and the environment (e.g.,
\cite{Mo96,sheth99}); recently, the topic has received a renewed interest in
the context of correlated excursion set theory (see
\cite{Faltenbacher10,ma11,paranjape14,zhang14}).

The (Eulerian) halo bias is defined as
\begin{equation}
b = 1+{1\over \delta_{c0}}\,\left[{N(M, \delta_c\rightarrow
M_0,\delta_{c0})\over N(M,\delta_c)\, V}-1\right]~;
\end{equation}
in the expression above what matters is the number of halos at redshift $z$
with density contrast $\delta_c$ that will end up in an environment with
volume $V$, mass $M_0\simeq \bar\rho V\gg M$, and density contrast
$\delta_{c0}\ll \delta_c$.

We will work under the peak-background split assumption, that corresponds to
consider conditions where $\Delta S\simeq S\gg S_0$. Then the detailed
computation of the conditional distribution appearing in eq. (3.13) is not
really needed, since its shape is close to the unconditional distribution
when written in terms of the appropriate scaling variable $\nu_c\equiv \Delta
\delta_c/ \sqrt{\Delta S}$. This is because the barrier of the conditional
problem $B(S,\delta_c)-B(S_0, \delta_{c0})\simeq B(S, \Delta \delta_c)$ tends
to the unconditional one as $S\gg S_0$. On considering that $\nu_c\simeq \nu
(1-\delta_{c0}/\delta_c)$ and $f(\nu_c)\simeq
f(\nu)-(\delta_{c0}/\delta_c)\,\nu f'(\nu)$ in the relevant limits $S\gg S_0$
and $\delta_c\gg \delta_{c0}$, one obtains
\begin{equation}
b(\nu)= 1+{1\over \delta_{c0}}\,\left[{\nu_c\, f(\nu_c)\over \nu\,
f(\nu)}-1\right]\simeq 1-{1\over \delta_c}\,\left[1+\,{{\rm d}\log f\over
{\rm d}\log\nu}\right]~.
\end{equation}

Using for simplicity the expression eq. (3.12) for the square-root barrier
with scatter we obtain
\begin{eqnarray}
\nonumber b(\nu) = 1 &+& {1\over \delta_c}\,\left\{{\nu^2+\beta\nu\over
1+\Sigma_\beta^2}-1+{1\over
1+\sqrt{\pi/2}\,\Gamma\nu\,e^{\Gamma^2\nu^2/2}\,{\rm
erfc}[-\Gamma\nu/\sqrt{2}]}+\right.\\
\\
\nonumber & - & \left. 3\,
{(\gamma\nu)^3+2\,\tau\,(\gamma\nu)^2+(1-2\,\gamma^2+\tau^2)\,\gamma\nu\over
(\gamma\nu)^3+3\,\tau\,(\gamma\nu)^2+3\,(1-2\,\gamma^2+\tau^2)\,\gamma\nu+(1-\gamma^2)
\,(1+2\,\beta)}-\left|{{\rm d}\log V/V_\star\over {\rm d}\log\nu}\right|\right\}~.
\end{eqnarray}
Note that the last term is due to the $\nu-$dependence of the quantity
$V/V_\star=(R/R_\star)^3$; it can be taken at an almost constant value
$\approx 0.2$ for the tophat filtering (cf. appendix A).

In figure 5 we show the results for uncorrelated excursion set (yellow line),
for the upcrossing approximation to correlated excursion set (red line), and
for our excursion set peak with scatter (orange line) from eq. (3.15). The
latter reproduces fairly well the outcomes from cosmological $N-$body
simulations by \cite{tinker10} (crosses); as it can be seen from the
residuals plotted in the inset, the relative deviation amounts to $\Delta\log
\nu\,f(\nu)\approx 0.05$ for $-1.5\lesssim \log\nu^2\lesssim 1.5$. Such
findings highlight that the peak constraint is a fundamental ingredient in
understanding the halo bias.

\section{Summary}

The formation and evolution of DM halos constitutes a complex issue, whose
attack ultimately requires cosmological $N-$body simulations on
supercomputers.

However, some analytic modeling is most welcome to better interpret the
simulation outcomes, to provide flexible analytic representations of the
results, to develop strategies for future setups, and to quickly explore the
effects of modifying the excursion set assumptions or the cosmological
framework.

In this vein, we have derived approximated, yet very accurate, analytical
expressions for the abundance and clustering of dark matter halos in the
excursion set peak framework. The latter is based on the standard excursion
set approach, but also includes the effects of a realistic filtering of the
density field, of a mass-dependent threshold for collapse, and of the
prescription from peak theory that halos tend to form around density maxima.

Our approximations works very well in reproducing the exact expressions for
different power spectra, collapse barrier and density filters. When adopting
a cold dark matter power spectra, a tophat filtering and a nonlinear
(square-root) barrier with scatter, our approximated halo mass function and
bias represent very well the outcomes of cosmological $N-$body simulations.

\appendix

\section{Excursion set peak parameters}

In this appendix we give some details on the computation of the parameters
$\Gamma$, $\gamma$ and $V/V_\star$ appearing in the main text.

The parameter $\Gamma$ is defined as
\begin{equation}
\Gamma\equiv {1\over\sqrt{4\,S\,\langle\dot\delta^2\rangle-1}}~;
\end{equation}
and relatedly $\gamma\equiv \Gamma/\sqrt{\Gamma^2+1}$. To give an explicit
expression for $\Gamma$, it is convenient to start from eqs. (2.1) and (2.5)
that define the quantities $S=\langle\delta_R^2\rangle$ and $C\equiv
\langle\delta_R\delta_{R'}\rangle$ in terms of the power spectrum $P(k)$ and
of the Fourier transform of the filter function $W_R(k)$. One has
\begin{eqnarray}
\nonumber\partial_{R} S &=& {1\over 2\pi^2}\int_0^\infty{\rm d}k\, k^{2}\, P(k)
\partial_{R} W_R^2(k)\\
&&\nonumber\\
\partial_{R'}C_{|R'=R} &=& {1\over 2\pi^2}\int_0^\infty{\rm d}k\, k^{2}\, P(k)
W_R(k)\,\partial_{R} W_R(k)\\
&&\nonumber\\
\nonumber\partial^2_{R,R'}C_{|R'=R} &=& {1\over 2\pi^2}\int_0^\infty{\rm d}k\,
k^{2}\, P(k) [\partial_{R} W_R(k)]^2
\end{eqnarray}
Combining these expressions yields $\langle\delta \dot\delta\rangle
=\partial_{R'}C_{|R'=R}/\partial_{R}S={1/2}$, and $\langle\dot
\delta^2\rangle = \partial^2_{R,R'}C_{|R'=R}/(\partial_R S)^2$; finally, one
finds
\begin{equation}
\Gamma = \left\{{\int_0^{\infty}{\rm
d}k~k^{2}\,P(k)\,W_R^2(k)\, \int_0^{\infty}{\rm d}k~k^{2}\,P(k)\, [\partial_R
W_R(k)]^2\over [\int_0^{\infty}{\rm
d}k~k^{2}\,P(k)\, W_R(k)\,\partial_R W_R(k)]^2}-1\right\}^{-1/2}~.
\end{equation}

The volume ratio $V/V_\star$ can be defined as $V/V_\star=(R/R_\star)^3$ in
terms of the characteristic radius
\begin{equation}
R_\star\equiv \left(3\,{R\,\partial_{R'}C_{|R'=R}\over
\partial^2_{R,R'}C_{|R'=R}}\right)^{1/2}= \left[3\, {\int_0^{\infty}{\rm
d}k~k^{2}\,P(k)\,W_R(k)\,R\,\partial_R W_R(k)\over \int_0^{\infty}{\rm
d}k~k^{2}\,P(k)\,[\partial_R W_R(k)]^2}\right]^{1/2}~.
\end{equation}

Next we detail the computations for a cold DM, and for a scale-invariant
power spectrum.

\subsection{Scale-invariant power spectrum}

When adopting a scale-invariant power spectrum $P(k)\propto k^n$, it is
convenient to work in terms of the variable $x=kR$ since $W_R(k)=W(x)$ for
both the Gaussian and tophat filters; then $\partial_R W_R(k)= k\, W'(x)$
holds, where the prime denotes derivative with respect to $x$.

For the Gaussian filter, one has
\begin{equation}
W_R(k)=e^{-x^2/2}~.
\end{equation}
Recalling the definition of the Euler Gamma function
\begin{equation}
\Gamma_{\rm E}(t)=2\,\int_0^\infty{\rm d}x~x^{2t-1}\,e^{-x^2}~
\end{equation}
and its basic property $\Gamma_{\rm E}(t+1)=t\,\Gamma_{\rm E}(t)$, from
eqs.~(A3) and (A4) it is easy to show that
\begin{equation}
\Gamma ={\Gamma_{\rm E}\left({n+5\over 2}\right)\over \Gamma_{\rm
E}\left({n+7\over 2}\right)\,\Gamma_{\rm E}\left({n+3\over
2}\right)-\Gamma_{\rm E}\left({n+5\over 2}\right)}=\sqrt{n+3\over 2}~.
\end{equation}
and that
\begin{equation}
{V\over V_\star} = \left[{\Gamma_{\rm E}\left({n+7\over
2}\right)\over 3\,\Gamma_{\rm E}\left({n+5\over
2}\right)}\right]^{3/2} = \left({n+5\over 6}\right)^{3/2}~.
\end{equation}

For the tophat filter, one has
\begin{equation}
W_R(k)={3\over x}\, j_1(x)~~~~~~~~~~W'(x)={3\over x}\, \left[j_0(x)-{3\over
x}\,j_1(x)\right]
\end{equation}
with $j_0(x)\equiv \sin x/x$ and $j_1(x)=(\sin x-x\cos x)/x^2$ the spherical
Bessel functions of the first kind. It is convenient to define the three
integrals
\begin{equation}
I_{00}\equiv \int_0^\infty{\rm d}x~x^{n+2}\,j_0^2(x)~~~~~~~~~ I_{01}\equiv
\int_0^\infty{\rm d}x~x^{n+1}\,j_0(x)\,j_1(x)~~~~~~~~~ I_{11}\equiv
\int_0^\infty{\rm d}x~x^{n}\,j_1^2(x)
\end{equation}
and using the properties of the Bessel functions to compute the ratios
\begin{equation}
{I_{00}\over I_{11}}={n\,(n-1)\,(n-3)\over
4\,(n+1)}~~~~~~~~~~~~~~~~~~{I_{01}\over I_{11}}={3-n\over 2}~.
\end{equation}
Exploiting eqs. (A3) and (A4) yields, after some tedious algebra,
\begin{equation}
\Gamma={|3\,I_{11}-I_{01}|\over
\sqrt{I_{11}\,I_{00}-I_{01}^2}}=\sqrt{(n+1)\,(n+3)\over n-3}~,
\end{equation}
and
\begin{equation}
{V\over V_\star} = \left(3\,{|3\,I_{11}-I_{01}|\over
{I_{00}+9\,I_{11}-6\,I_{01}}}\right)^{3/2} = \left[{n\,(n+5)\over
6\,(n+1)}\right]^{3/2}~.
\end{equation}
Note that to ensure convergence of the integrals in eqs.~(A10) and positivity
of $\Gamma^2$ and $(V/V_\star)^2$ it is required that the spectral index
falls in the range $-3<n<-1$.

\subsection{Cold DM power spectrum}

For the cold DM power spectrum, both $\Gamma$ (or
$\gamma=\Gamma/\sqrt{1+\Gamma^2}$) and $V/V_\star$ depend on scale. In figure
6 we plot these quantities as a function of the scaling variable $\nu$ for
both the Gaussian and tophat filters.

As to $\Gamma$, the dependence on $\nu^2$ can be represented to better than
$5\%$ in the range $-1.5\lesssim \log \nu^2\lesssim 1.5$ by the following
expression:
\begin{equation}
\log{\Gamma} \simeq \Gamma_0 + \Gamma_1\, \log \nu^2 + \Gamma_2\,(\log
\nu^2)^2~;
\end{equation}
the fitting parameters $(\Gamma_0,\Gamma_1,\Gamma_2)$ amount to
$(-0.071,0.119,-0.019)$ for the Gaussian and $(-0.291,0.086,-0.030)$ for the
tophat filtering. The dependence on $\nu$ is mild (almost logarithmic), and
can be neglected to a first approximation in the excursion set expressions;
taking the values $\Gamma\approx 0.85$ (or $\gamma\approx 0.65$) and $0.51$
(or $\gamma\approx 0.46$) corresponding to $\nu\approx 1$ for the Gaussian
and tophat filtering works pretty well.

As to $V/V_\star$, the following expression applies:
\begin{equation}
\log{V/V_\star} \simeq V_0 + V_1\, \log \nu^2 + V_2\,(\log \nu^2)^2~,
\end{equation}
with parameters $(V_0,V_1,V_2)$ amounting to $(-0.459,0.121,0.009)$ for the
Gaussian and to $(-0.179,0.172,0.036)$ for the tophat filtering.

\acknowledgments Work supported in part by INAF and MIUR. We thank the
anonymous referee for helpful and constructive comments. We acknowledge
stimulating discussions with A. Cavaliere, P.S. Corasaniti, and P. Salucci.
A.L. is grateful to SISSA for warm hospitality.

\clearpage
\begin{figure*}
\centering
\includegraphics[width=14cm]{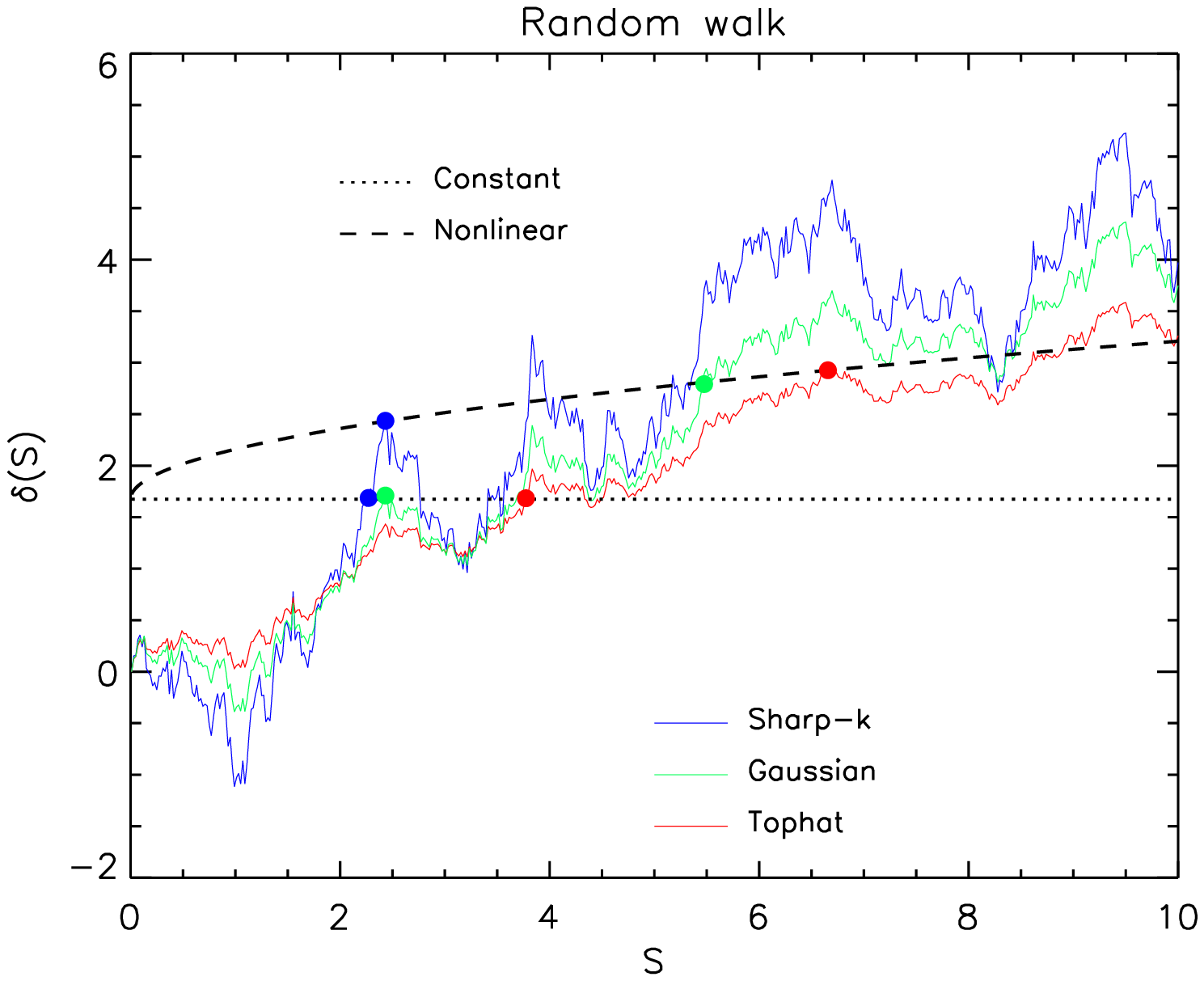}
\caption{Examples of random walks executed by the overdensity contrast
$\delta_R$ as a function of the variance $S_R$, when a sharp $k$-space (blue
line), a Gaussian (green line) or a tophat (red line) smoothing filter is
adopted. The dotted and dashed lines illustrate the constant and square-root
barrier at redshifts $z = 0$; the dots illustrate the location of first
crossing. It is easily seen that the random walks for the gaussian and tophat
filters vary less erratically than that for the sharp $k-$space, as a
consequence of the correlations between the steps.}
\end{figure*}

\clearpage
\begin{figure*}
\centering
\includegraphics[width=12cm]{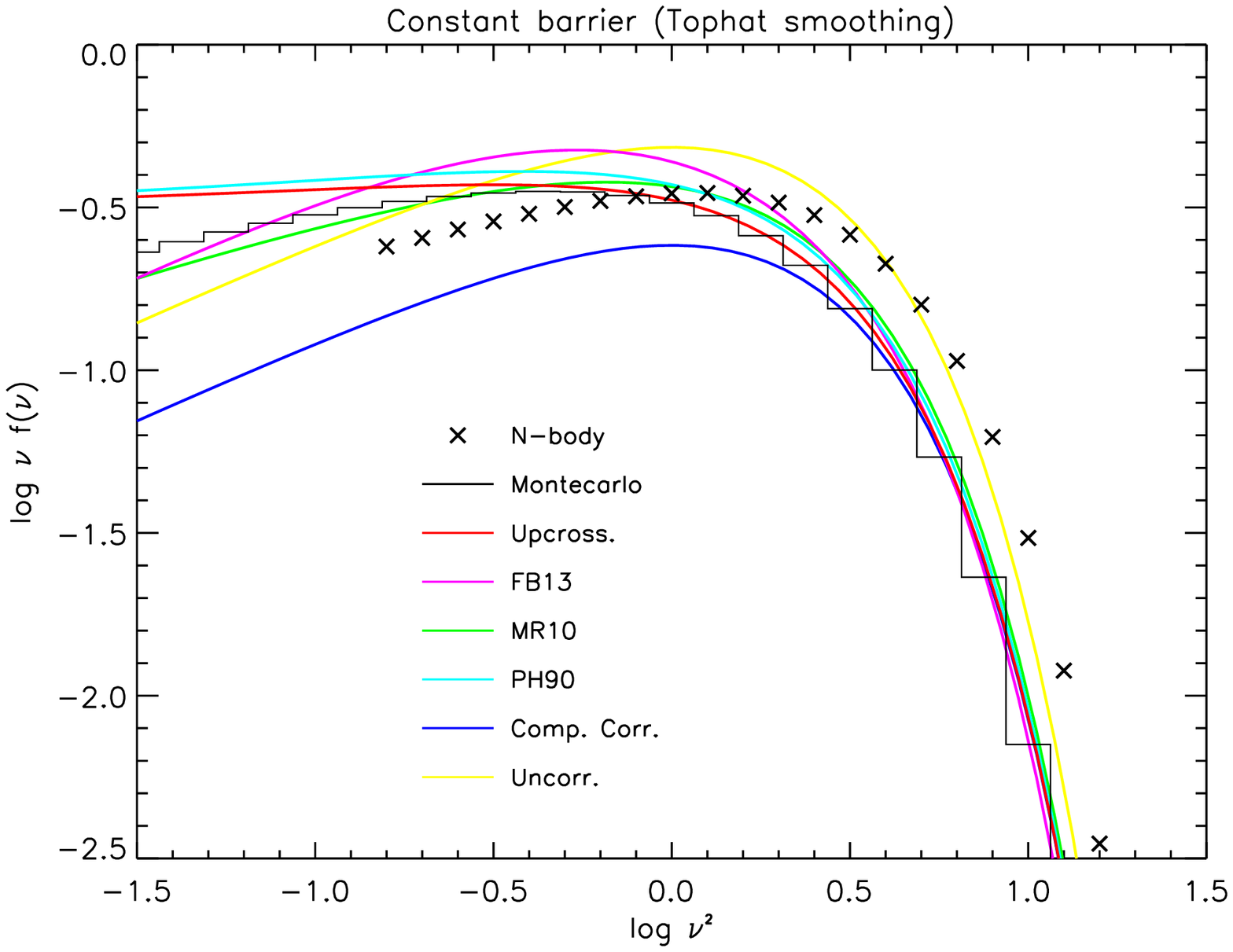}\\\includegraphics[width=12cm]{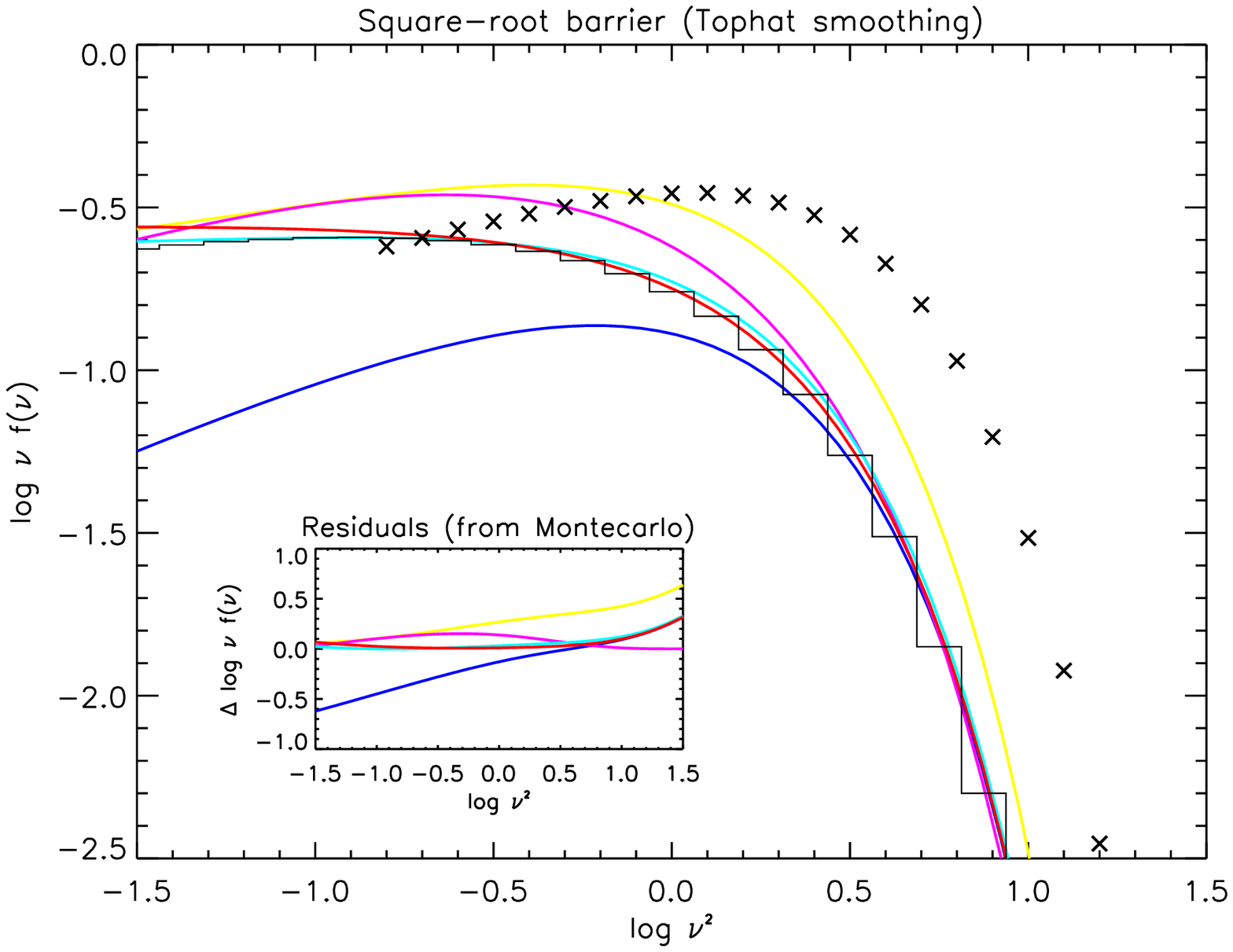}
\caption{First crossing distribution for a tophat smoothing filter, adopting
a constant (top panel) or a square-root (bottom panel) barrier after eq.
(2.2) and the standard excursion set theory prescription. The black histogram
refers to the exact solution from Montecarlo simulations, red line to the
upcrossing approximation by \cite{bond91,musso12}, magenta line to the
approximation by \cite{farahi13}, green line to the approximation by
\cite{maggiore10}, cyan line to the approximation by \cite{peacock90}, blue
line the completely correlated limit by \cite{paranjape12}, and yellow line
to the result for a sharp $k-$space. Crosses illustrate the outcomes of
cosmological $N-$body simulations from \cite{tinker08}. In the bottom panel,
the inset shows the residuals with respect to the Montecarlo outcome.}
\end{figure*}

\clearpage
\begin{figure*}[!t]
\centering
\includegraphics[width=12cm]{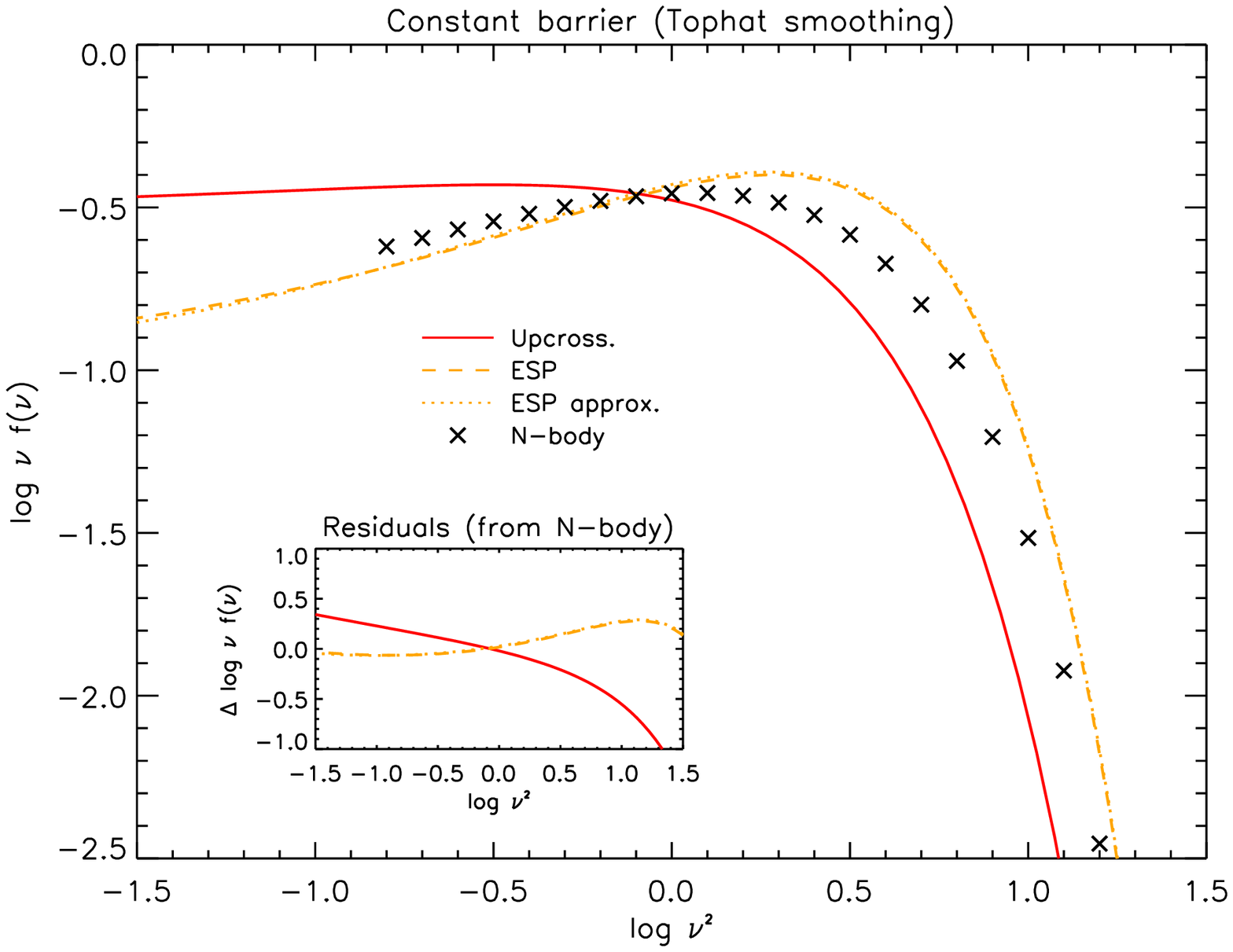}\\\includegraphics[width=12cm]{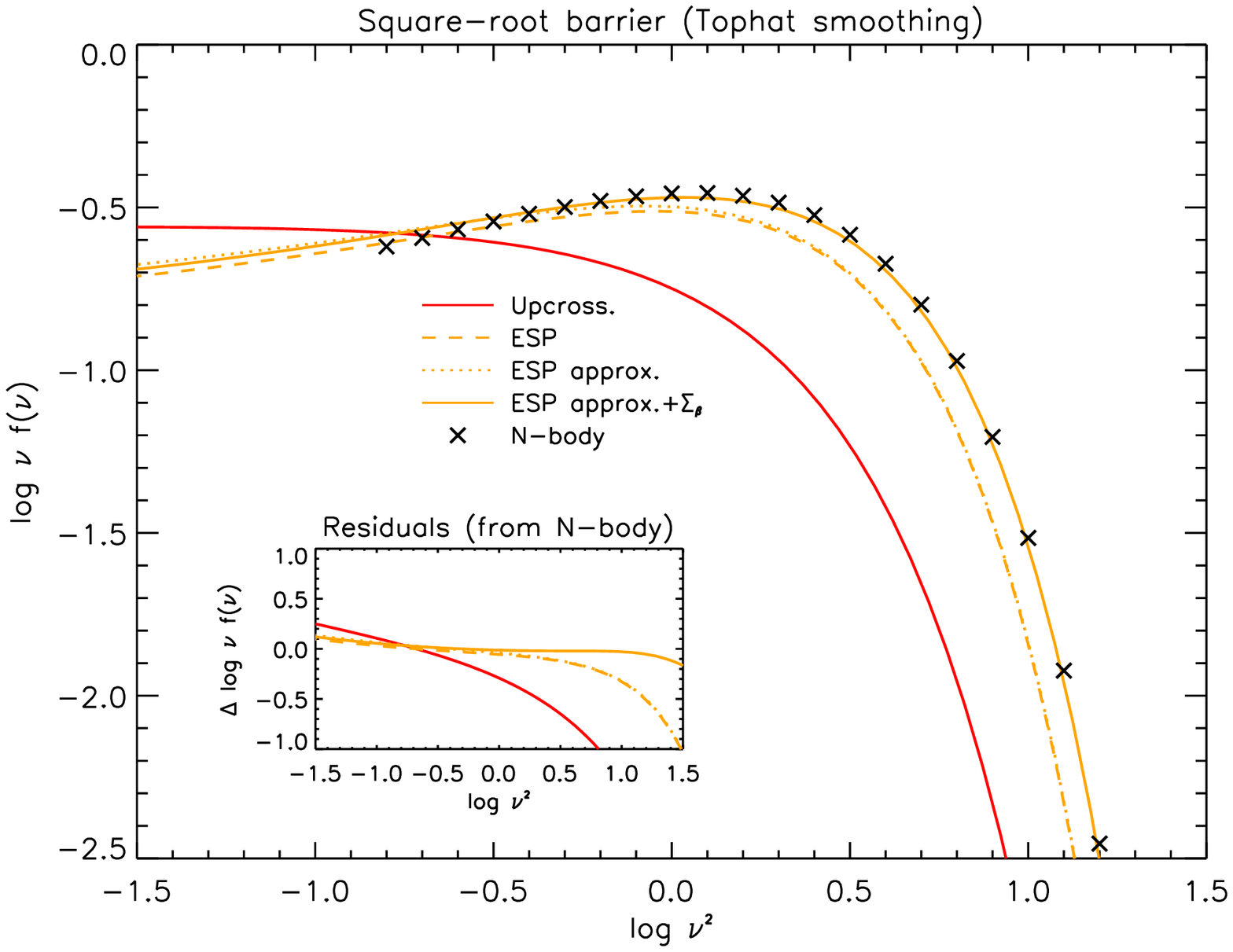}
\caption{First crossing distribution for a tophat smoothing filter, adopting
a constant (top panel) or a square-root (bottom panel) barrier and the
excursion set peak prescription. Red line is the result for the standard
excursion set approach (actually the upcrossing approximation, as in figure
2), dashed orange line refers to the exact result for the excursion set peak
(cf. eq. 2.9), dotted line is our approximation of eqs. (3.10) and (3.11),
and solid line in the bottom panel includes the scatter on the barrier after
eq. (3.12). Crosses illustrate the outcomes of cosmological $N-$body
simulations from \cite{tinker08}. In both panels the inset shows the
residuals with respect to the $N-$body outcome.}
\end{figure*}

\clearpage
\begin{figure*}[!t]
\centering
\includegraphics[width=12cm]{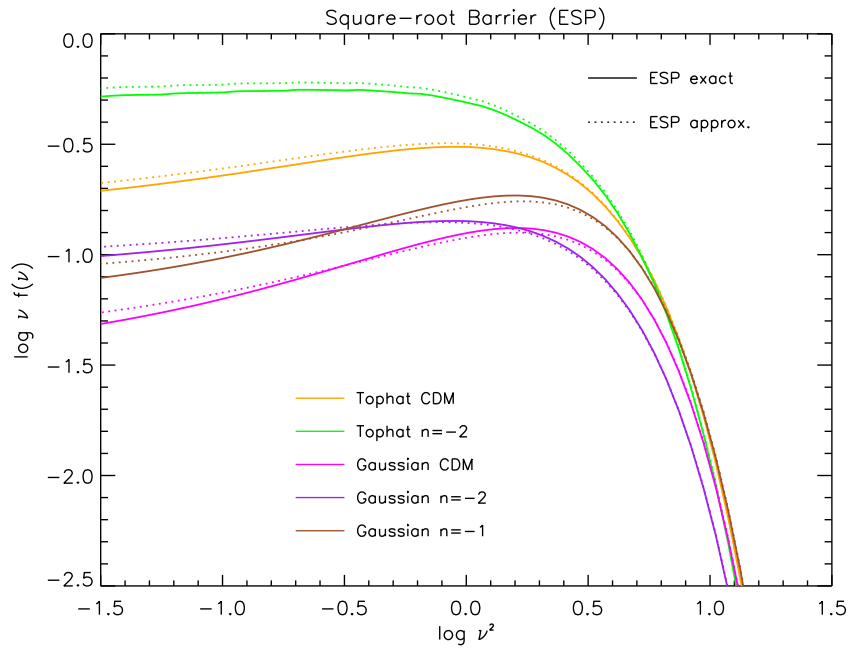}
\caption{First crossing distribution for a tophat smoothing filter, adopting
a square-root barrier after eq. (2.2) and the excursion set peak
prescriptions. Solid lines illustrate the exact result after eq. (2.9), and
dashed line is our approximation of eq. (3.11). As detailed in the legend,
orange and green lines refer to a tophat smoothing filter and the standard
cold DM or a scale invariant power-spectrum with index $n=-2$; magenta,
purple and brown lines refer to a Gaussian smoothing filter and a cold DM or
scale-invariant power spectrum with index $n=-2$ and $n=-1$.}
\end{figure*}

\clearpage
\begin{figure*}[!t]
\centering
\includegraphics[width=12cm]{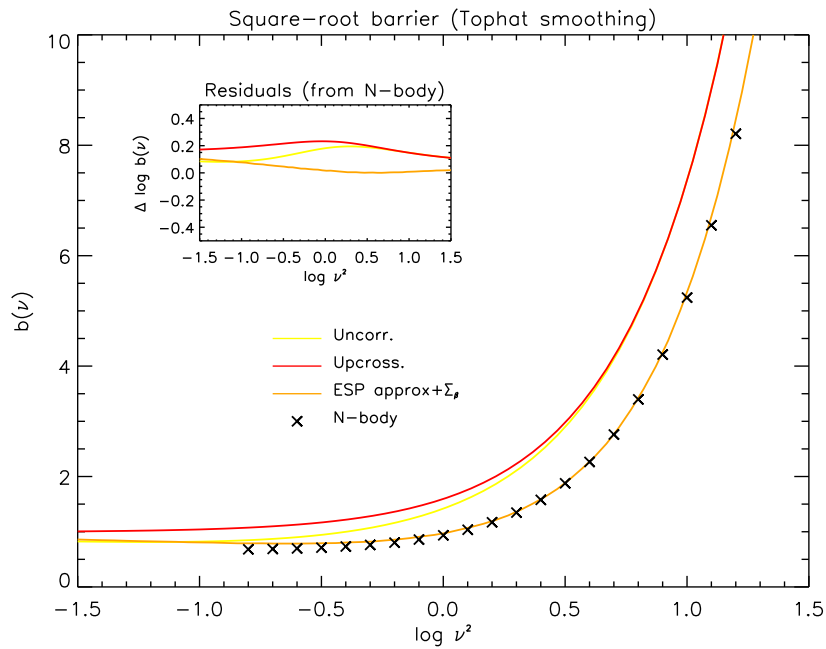}
\caption{Halo bias in the background-split approximation, for a tophat
smoothing filter and a square-root barrier. Yellow line refers to
uncorrelated (Markovian) walks, red line to the upcrossing approximation for
correlated excursion set theory, and orange line to our approximation for the
excursion set peak of eqs. (3.14) and (3.15). Crosses illustrate the outcomes
of cosmological $N-$body simulations from \cite{tinker10}. The inset shows the
residuals with respect to the $N-$body outcome.}
\end{figure*}

\clearpage
\begin{figure*}[!t]
\centering
\includegraphics[width=12cm]{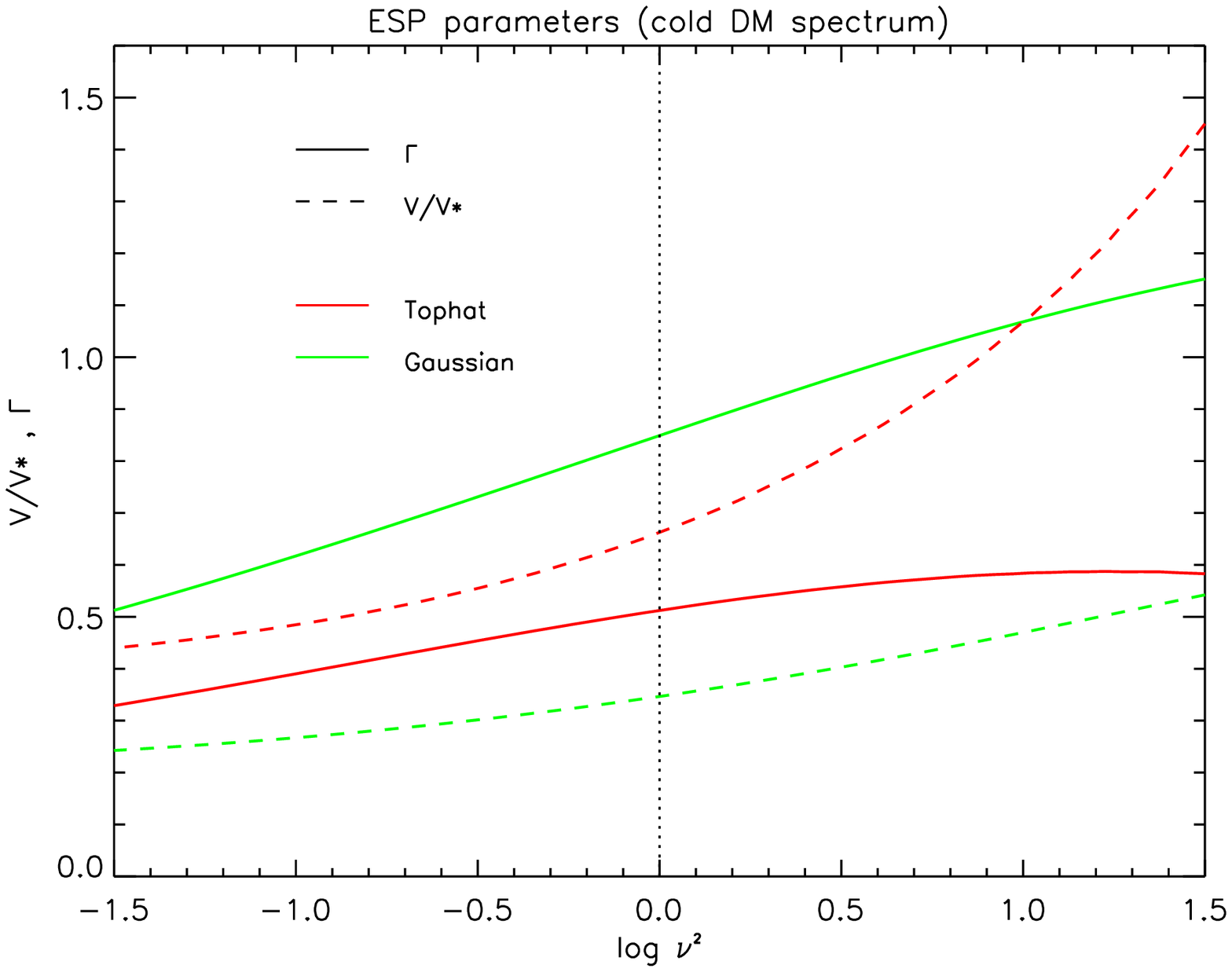}
\caption{Excursion set peak parameters $\Gamma$ (solid lines) and $V/V_\star$
(dashed lines) as a function of the scaling variable $\nu$, for Gaussian
(green) and tophat (red) filtering, cf. appendix A. The vertical dotted
line marks the location $\nu=1$.}
\end{figure*}


\begin{thebibliography}{99}

\bibitem{sheth01}R.K. Sheth, H.J. Mo, and G. Tormen, \emph{Ellipsoidal
    collapse and an improved model for the number and spatial distribution of
    dark matter haloes}, \emph{MNRAS} \textbf{323} (2001) 1

\bibitem{springel05}V. Springel, et al., \emph{Simulations of the
    formation, evolution and clustering of galaxies and quasars},
    \emph{Nature} \textbf{435} (2005) 629

\bibitem{warren06}M.S. Warren, K. Abazajian, D.E. Holz, and L. Teodoro,
    \emph{Precision Determination of the Mass Function of Dark Matter Halos},
    \emph{ApJ} \textbf{646} (2006) 881

\bibitem{reed07}D.S. Reed, R. Bower, C.S. Frenk, A. Jenkins, and T. Theuns,
    \emph{The Halo Mass Function into the Dark Ages}, \emph{MNRAS} \textbf{374} (2007) 2

\bibitem{tinker08}J. L. Tinker, et al., \emph{Toward a Halo Mass Function for
    Precision Cosmology: The Limits of Universality}, \emph{ApJ} \textbf{688} (2008) 709

\bibitem{murray13}S.G. Murray, C. Power, and A.S.G. Robotham, \emph{How well
    do we know the halo mass function?}, \emph{MNRAS} \textbf{434} (2013) L61

\bibitem{watson13}W.A. Watson, I.T. Iliev, A. D'Aloisio, A. Knebe, P.R.
    Shapiro, and G. Yepes, \emph{The halo mass function through the cosmic ages},
    \emph{MNRAS} \textbf{433} (2013) 1230

\bibitem{bond91}J.R. Bond, S. Cole, G. Efstathiou, and N. Kaiser,
    \emph{Excursion set mass functions for hierarchical gaussian fluctuations},
    \emph{ApJ} \textbf{379} (1991) 440

\bibitem{corasaniti11}P.S. Corasaniti, and I. Achitouv, \emph{Toward a
    Universal Formulation of the Halo Mass Function}, \emph{Phys Rev. L.}
    \textbf{106} (2011) 1302

\bibitem{epstein83}R.I. Epstein, \emph{Proto-galactic perturbations},
    \emph{MNRAS} \textbf{205} (1983) 207

\bibitem{lapi13}A. Lapi, P. Salucci, and L. Danese, \emph{Statistics of
    dark matter halos from the excursion set approach}, \emph{ApJ}
    \textbf{772} (2013) 85

\bibitem{maggiore10}M. Maggiore, and A. Riotto, \emph{The halo mass
    function from excursion set theory. I. Gaussian fluctuations with
    non-markovian dependence on the smoothing scale}, \emph{ApJ} \textbf{711} (2010) 907

\bibitem{musso12}M. Musso, and R.K. Sheth, \emph{One step beyond: the
    excursion set approach with correlated steps}, \emph{MNRAS}
    \textbf{423} (2012) L102

\bibitem{paranjape12}A. Paranjape, T.-Y. Lam, and R.K. Sheth, \emph{Halo
    abundances and counts-in-cells: the excursion set approach with
    correlated steps}, \emph{MNRAS} \textbf{420} (2012) 1429

\bibitem{press74}W.H. Press, and P. Schechter, \emph{Formation of Galaxies
    and Clusters of Galaxies by Self-Similar Gravitational Condensation},
    \emph{ApJ} \textbf{187} (1974) 425

\bibitem{sheth02}R.K. Sheth, and G. Tormen, \emph{An excursion set model of
    hierarchical clustering: ellipsoidal collapse and the moving
    barrier}, \emph{MNRAS} \textbf{329} (2002) 61

\bibitem{appel90}L. Appel, and B.J.T. Jones, \emph{The Mass Function in
    Biased Galaxy Formation Scenarios}, \emph{MNRAS} \textbf{245} (1990) 522

\bibitem{bardeen86}J.M. Bardeen, J.R. Bond, N. Kaiser, and A.S. Szalay,
    \emph{The statistics of peaks of Gaussian random fields},
    \emph{ApJ} \textbf{304} (1986) 15

\bibitem{bond96}J. Bond, and S. Myers, \emph{The Peak-Patch Picture of Cosmic
    Catalogs. I. Algorithms}, \emph{ApJS} \textbf{103} (1996) 1

\bibitem{manrique98}A. Manrique, A. Raig, J.M. Solanes, G. Gonzalez-Casado,
    P. Stein, and E. Salvador-Sole, \emph{The Effects of the Peak-Peak
    Correlation on the Peak Model of Hierarchical Clustering}, \emph{ApJ} \textbf{499} (1998) 548

\bibitem{jedamzik95}K. Jedamzik, \emph{The cloud-in-cloud problem in the
    press-schechter formalism of hierarchical structure formation},
    \emph{ApJ} \textbf{448} (1995) 1

\bibitem{nagashima01}M. Nagashima, \emph{A Solution to the Missing Link in
    the Press-Schechter Formalism}, \emph{ApJ}, \textbf{562} (2001) 7

\bibitem{paranjape13a}A. Paranjape, R.K. Sheth, and V. Desjacques,
    \emph{Excursion set peaks: a self-consistent model of dark halo
    abundances and clustering}, \emph{MNRAS} \textbf{431} (2013) 1503

\bibitem{planck13}\textsl{Planck} Collaboration 2013, \emph{Planck 2013
    results. XVI. Cosmological parameters}, \emph{A\&A}, in press [preprint arXiv:1303.5076]

\bibitem{sugiyama95}N. Sugiyama,\emph{Cosmic Background Anisotropies in Cold
    Dark Matter Cosmology }, \emph{ApJS} \textbf{100} (1995) 281

\bibitem{eke96}V.R. Eke, S. Cole, and C.S. Frenk, \emph{Cluster evolution as
    a diagnostic for Omega}, \emph{MNRAS} \textbf{282} 263

\bibitem{robertson09}B.E. Robertson, A.V. Kravtsov, J. Tinker, and
    A.R. Zentner, \emph{Collapse barriers and halo abundance: testing
    the excursion set ansatz}, \emph{ApJ} \textbf{696} (2009) 636

\bibitem{despali13}G. Despali, G. Tormen, and R.K. Sheth, \emph{Ellipsoidal
    halo finders and implications for models of triaxial halo formation},
    \emph{MNRAS} \textbf{431} (2013) 1143

\bibitem{benson13}A.J. Benson, et al., \emph{Dark matter halo merger
    histories beyond cold dark matter - I. Methods and application to warm
    dark matter}, \emph{MNRAS} \textbf{428} (2013) 1774

\bibitem{zentner07}A.R. Zentner, \emph{The Excursion Set Theory of Halo Mass
    Functions, Halo Clustering, and Halo Growth}, \emph{Journ. Mod. Phys.}
    \textbf{16} (2007) 763

\bibitem{zhang06}J. Zhang, and L. Hui, \emph{On random walks with a general
    moving barrier}, \emph{ApJ} \textbf{641} (2006) 641

\bibitem{farahi13}A. Farahi, and A. Benson, \emph{Excursion set theory for
    correlated random walks}, \emph{MNRAS} \textbf{433} (2013) 3428

\bibitem{peacock90}J.A. Peacock, and A.F. Heavens, \emph{Alternatives to
    the Press and Schechter cosmological mass function}, \emph{MNRAS}
    \textbf{243} (1990) 133

\bibitem{paranjape13b}A. Paranjape, and R.K. Sheth, \emph{Peaks theory and
    the excursion set approach}, \emph{MNRAS} \textbf{426} (2013) 2789

\bibitem{Mo96}H.J. Mo, and S.D.M. White, \emph{An analytic model for the
    spatial clustering of dark matter haloes}, \emph{MNRAS} \textbf{282}
    (1996) 347

\bibitem{sheth99}R.K. Sheth, and G. Tormen, \emph{Large-scale bias and the
    peak background split}, \emph{MNRAS} \textbf{308} (1999) 119

\bibitem{Faltenbacher10}A. Faltenbacher, and S.D.M. White, \emph{Assembly
    bias and the dynamical structure of dark matter halos},
    \emph{ApJ} \textbf{708} (2010) 469

\bibitem{ma11}C.-P. Ma, M. Maggiore, A. Riotto, and J. Zhang, \emph{The bias
    and mass function of dark matter haloes in non-Markovian extension of the
    excursion set theory}, \emph{MNRAS} \textbf{411} (2011) 2644

\bibitem{paranjape14}A. Paranjape, E. Sefusatti, K. C. Chan, V. Desjacques,
    P. Monaco, R.K. Sheth, \emph{Bias deconstructed: unravelling the scale
    dependence of halo bias using real-space measurements},
    \emph{MNRAS} \textbf{436} (2013) 449

\bibitem{zhang14}J. Zhang, C.-P. Ma, and A. Riotto, \emph{Dark-matter halo
    assembly bias: environmental dependence in the non-markovian excursion-set
    theory}, \emph{ApJ} \textbf{782} (2014) 44

\bibitem{tinker10}J.L. Tinker, et al., \emph{The large-scale bias of dark
    matter halos: numerical calibration and model tests},
    \emph{ApJ} \textbf{724} (2010) 878

\end{thebibliography}
\end{document}